\documentclass[
    aps,
    10pt,
    prd,
    notitlepage,
    twocolumn,
    tightenlines,
    nofootinbib]{revtex4-1}

\usepackage[linktocpage,breaklinks]{hyperref}
\usepackage[usenames,dvipsnames]{xcolor}

\usepackage{aas_macros}
\usepackage{amsmath}
\usepackage{amssymb}
\usepackage{amsfonts}
\usepackage{txfonts}
\usepackage{bm}
\usepackage{stmaryrd}
\usepackage{tensor}
\usepackage[utf8]{inputenc}

\usepackage{graphicx}
\usepackage{epsfig}
\usepackage{epstopdf}

\usepackage{natbib}
\usepackage{hyperref}
\usepackage{cleveref}

\definecolor{mred}{RGB}{127,0,25}
\definecolor{mdgr}{RGB}{51,51,51}
\definecolor{mag}{RGB}{211, 54, 130}
\definecolor{verm}{RGB}{164, 25, 0}

\hypersetup{colorlinks=true,
            citecolor=NavyBlue,
            linkcolor=NavyBlue,
            urlcolor=NavyBlue}

\newcommand{\msun}{$_{\odot}$}

\begin{document}
\title{Neutron star pulse profile observations as extreme gravity probes}
\begin{abstract}

The x-ray emission of hot spots on the surface of neutron stars is the
prime target of the \emph{Neutron star Interior Composition Explorer} (NICER).
These x-ray pulse profiles not only encode information of the bulk properties of these stars,
which teaches us about matter at supranuclear densities, but also about the spacetime
curvature around them which teaches us about relativistic gravity.
We explore the possibility of performing strong-gravity tests with NICER observations
using a recently developed pulse profile model beyond general relativity.
Our results suggest that NICER can in principle place constraints on
deviations from general relativity due to an additional scalar degree of
freedom which are independent and competitive relative to constraints with
binary pulsar observations.

\end{abstract}

\author{Hector O. Silva}
\email{hector.okadadasilva@montana.edu}
\affiliation{eXtreme Gravity Institute,
Department of Physics, Montana State University, Bozeman, Montana 59717 USA}

\author{Nicol\'as Yunes}
\email{nicolas.yunes@montana.edu}
\affiliation{eXtreme Gravity Institute,
Department of Physics, Montana State University, Bozeman, Montana 59717 USA}

\date{\today}

\maketitle

%%%%%%%%%%%%%%%%%%%%%%%%%%%%%%%%%%%%%%%%%%%%%%%%%%%%%%%%%%%%%%%%%%%%%%
\section{Introduction}
Extreme and mysterious, neutron stars are an
unavoidable consequence of general relativity at high enough
densities. With typical masses between 1 and 2 M$_\odot$,
but radii of only $\sim 11$ km, their energy densities
can exceed nuclear saturation in their inner core. When this occurs,
matter can transmute into exotic forms that are impossible to
replicate in laboratories on Earth, and understanding the physics
of such extreme matter remains an open problem in
nuclear astrophysics.

Neutron stars are not just a laboratory for nuclear physics.
Their high densities also imply strong gravitational fields that exceed
those that can be probed in the Solar System by seven orders of
magnitude. Understanding the physics of such strong gravity objects
requires the use of a relativistic theory of gravity,
like general relativity~\cite{Oppenheimer:1939ne}.
Neutron star observations can therefore provide invaluable
clues about both nuclear astrophysics and
relativistic gravity, as was achieved with the recent observation
of gravitational waves from the merger of two neutron stars~\cite{TheLIGOScientific:2017qsa,Soares-Santos:2017lru,Abbott:2018exr}.

We here report on the first results of a new program to systematically
study how strong-field gravity can be probed with a particular set of neutron
star observations: pulse profiles emitted by radiating hot spots
on the star's surface~\cite{Arzoumanian:2009qn}.
Neutron stars can have hot spots on their surface, i.e.~regions
where the surface temperature is much higher than the average, due to
the impact of accreted material that is pulled in from a nearby, less dense star
or by localized heating due to magnetospheric currents (see
e.g.~\cite{Poutanen:2008pg,Bogdanov:2008qm,Ozel:2012wu,Watts:2016uzu,Degenaar:2018lle}).
These hot spots rotate rapidly with the star, emitting x-rays that
trace the spacetime geometry as they leave the star, producing a pulse profile
upon detection.
NASA's \emph{Neutron star Interior Composition Explorer} (NICER)
is currently detecting these pulse profiles with unprecedented
(timing) resolution through rotation-resolved spectroscopy of thermal and
non-thermal emission~\cite{GendreauSPIE2012,ArzoumaninanSPIE2014,GendreauNature2017}.
Can NICER constrain (or reveal) deviations from general relativity
in the strong-field regime of neutron stars?

The constraining power of such observations depend
sensitively on our ability to accurately model the pulse profile within and
outside general relativity. In~\cite{Silva:2018yxz} we presented a
complete toolkit to model x-ray pulse profiles in an entire class of
well-motivated modifications to general relativity: scalar-tensor
gravity~\cite{Damour:1992we,Chiba:1997ms,Sotiriou:2015lxa}.
In this class, gravity is still described by the curvature of spacetime,
characterized by the spacetime metric as in general relativity,
but the metric is influenced by a scalar field that can be excited
in sufficiently dense environments, as in neutron stars.
Since the scalar field is usually negligible in the Solar System, scalar-tensor
theories have survived a plethora of experimental tests~\cite{Will:2014kxa},
while remaining prime candidates for tests with
neutron stars~\cite{Berti:2015itd,Kramer:2016kwa,Doneva:2017jop}.

The toolkit is fast, computationally efficient, covers a generic
family of scalar-tensor theories of gravity and it is ready to use for
probing strong gravity with NICER observations. The resulting pulse profiles
include Doppler shifts, relativistic aberration and time-delay effects,
thus extending the work of~\cite{Miller:1997if} to scalar-tensor theories
and the work of~\cite{Sotani:2017rrt} to the required level of astrophysical
realism.
These effects are crucial for constructing sufficiently accurate
pulse-profile models, thus enabling, for the first time, a serious data
analysis investigation of the strength with which NICER observations
can probe strong gravity.
As a first step toward such an investigation,
we present an approximate, restricted Bayesian
calculation and find that a NICER observation consistent with general
relativity could allow for constraints on scalar tensor theories that can be
comparable to the most stringent constraints from binary pulsar
observations to date~\cite{Freire:2012mg,Shao:2017gwu,Anderson:2019eay,Archibald:2018oxs}.

%%%%%%%%%%%%%%%%%%%%%%%%%%%%%%%%%%%%%%%%%%%%%%%%%%%%%%%%%%%%%%%%%%%%%%
\section{Light curve modeling in scalar-tensor gravity}

The x-ray photons emitted by hot spots at the surface of a neutron star trace
its exterior spacetime geometry. If the star is isolated, or if its companion
is far away and we focus on the neutron star's vicinity, the exterior spacetime
can be well-approximated by the Schwarzschild spacetime, provided the star
rotates much slower than the Kepler limit; this is indeed the case for some
targets of NICER, which have spin frequencies below $300$ Hz.
In scalar-tensor gravity, the gravitational interaction is mediated by a
scalar field $\varphi$ in addition to the usual metric tensor $g_{\mu\nu}$ of
general relativity.
Here we consider a wide class of theories in which a massless scalar field is
minimally coupled to gravity, yet matter fields are coupled to the product of the
function $A(\varphi)$ and the metric tensor $g_{\mu\nu}$. The function $A$ determines the scalar-tensor model
under consideration.
The presence of the scalar field modifies the exterior spacetime of a static,
spherically symmetric star which is no longer the usual Schwarzschild spacetime
of general relativity, but instead is the Just
spacetime~\cite{Just1959,Damour:1992we}. This spacetime depends not only on the
baryonic mass of the star $m$, but also on the ``charge'' or field strength $q$ of the
scalar field $\varphi$.
The modifications relative to the Schwarzschild spacetime are
controlled by the scalar-charge-to-mass ratio
$Q \equiv q c^2 / (G m)$~\cite{Just1959,Damour:1992we}, which is zero
in general relativity.
See~\cite{Silva:2018yxz} for a detailed discussion.

\begin{figure}[t]
\includegraphics[width=0.48\textwidth]{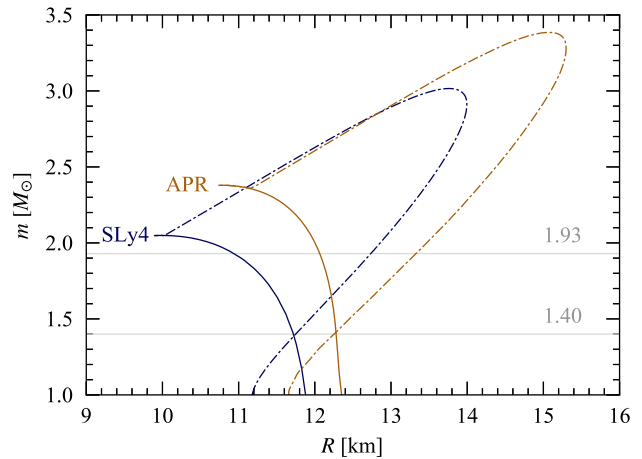}
\caption{Masses and radii of neutron stars in scalar-tensor gravity. The
solid curves represent families of stars with varying central density
in general relativity using the SLy4~\cite{Douchin:2001sv} and APR~\cite{Akmal:1998cf}
equations of state.
Neutron stars in scalar-tensor gravity with fixed $\alpha_0 = 10^{-5}$ and
$\beta_0 = -7$ are shown with dot-dashed curves, which bound the scalarized
solutions (at fixed $\alpha_0$) with increasing $|\beta_0|$.
The value $\beta_0 = -7$ is ruled-out by binary pulsar
observations, yet we include it here as an illustration of how large the
deviations from general realtivity can be in scalar-tensor theories which
admit spontaneous scalarization.
When $\beta_0$ is close to the scalarization threshold (i.e. $\approx -4.35$) the
differences between the general relativity and scalar-tensor gravity predictions
for the mass-radius curves are negligible independently of the equation of state.
The horizontal lines indicate that, given an EoS, neutron stars with larger
mass (or equivalently larger central densities) have greater variation in their
radius as $\beta_0$ becomes more negative.
}
\label{fig:mass_radius}
\end{figure}

As a concrete example of scalar-tensor gravity we consider the two-parameter
theory of~\cite{Damour:1993hw,Damour:1996ke} in which
$\ln A(\varphi) \equiv \alpha_0 \varphi + \beta_0 \varphi^2 / 2$.
The parameter $\alpha_0$ is related to the Brans-Dicke
parameter $\omega_{\rm BD}$ via $\alpha_{0}^2 = 1/(2 \omega_{\rm BD} + 3)$,
and it is stringently constrained to $\alpha_0 \lesssim 3.4 \times 10^{-3}$ by
Shapiro time-delay measurements of the Cassini spacecraft~\cite{Bertotti:2003rm}.
On the other hand, $\beta_0$ is unconstrained by Solar System experiments.
When $\beta_0 \gtrsim -4.35$, neutron stars have a small
$Q \propto \alpha_0$~\cite{Anderson:2019hio} and they are similar to those
of general relativity.
When $\beta_0 \lesssim -4.35$~\cite{Harada:1997mr,Novak:1998rk},
regardless of the equation of state (EoS)~\cite{Silva:2014fca},
the theory admits a new branch of solutions, where stars with
$Q \approx {\cal O}(1)$ are energetically favored over
their general relativity counterparts, due to a non-perturbative
effect called spontaneous scalarization\footnote{When the cosmological evolution of
the scalar field is taken into account, $\beta < 0$ violates current
Solar System constraints unless one fine-tunes the initial conditions
for the cosmological evolution. When $\beta_0 > 0$, these constraints
are satisfied~\cite{Damour:1992kf} at the cost of disallowing
scalarization, unless $\beta_0 \gtrsim 100$~\cite{Mendes:2014ufa}.
We here study this theory with $\beta < 0$ as a toy-model for non-perturbative
strong-gravity effects, generically predicted in gravitational
theories with extra degrees of freedom (see e.g.~\cite{Silva:2017uqg,Doneva:2017bvd,Annulli:2019fzq}),
and because it allows for comparison with binary pulsar constraints~\cite{Freire:2012mg,Anderson:2019eay}.
}~\cite{Damour:1993hw,Damour:1996ke}.
Numerically integrating the stellar structure equations in this theory,
we obtain $m$, $q$ and $A(\varphi)$, which completely determine the spacetime.
Figure~\ref{fig:mass_radius} shows that the scalar field
can significantly modify the stellar structure and thus the
spacetime around it.

With the spacetime geometry at hand, we can construct a
pulse profile model. In general relativity, the models used for
NICER data analysis are semi-analytic, as full null geodesic
ray tracing for the construction of millions of profiles in a Markov-Chain
Monte-Carlo exploration of the likelihood is computationally
prohibitive. A simple yet accurate model commonly used in general relativity is
Schwarzschild+Doppler~\cite{Miller:1997if,Poutanen:2003yd,Poutanen:2006hw},
whose generalization to scalar-tensor theory leads to the ``Just+Doppler"
model~\cite{Silva:2018yxz}, which includes Schwarzschild+Doppler
in the $Q=0$ limit.
As in the Schwarzschild+Doppler case, the Just+Doppler model
is constructed under some simplifications: the star is assumed spherical
(i.e.~deformations due to rotation are ignored) and the effects of frame-dragging
in the photon's geodesic are neglected.
As in general relativity, these effects in scalar-tensor theory are
likely to have a negligible influence on the resulting pulse profile, as long as
the star rotates slowly ($\lesssim 300$ Hz).
This is because for slowly-rotating stars
we do not expect much of a deformation away from sphericity due to rotation.
Moreover, the frame-dragging of inertial frames in scalar-tensor gravity
is very similar to that of general relativity, justifying extrapolating
the conclusions from one theory to the other~\cite{Sotani:2012eb}.

For our analysis, we further assume that the hot spot is small (relative to the
size of the star) and isotropically radiating.
In particular, we assume that the spot has a small angular radius $\Delta \theta$
and that it irradiates with a blackbody spectrum of constant $k_{\rm B} T$
measured in the frame comoving with the hot spot ($k_{\rm B}$ being
Boltzmann's constant), while the rest of the star is dark.
As often done in the literature, we consider a single hot spot (see e.g.~\cite{Lo:2013ava,Lo:2018hes})
and (for simplicity) neglect any other radiation coming from the background.
Finally, we assume that the observer collects photons in the
soft x-ray band in a single energy channel of $1$ keV.
Although these simplifications are too rough to be implemented directly
in an actual analysis of NICER data~\cite{Braje:2000qb,Cadeau:2004gm,Cadeau:2006dc,Morsink:2007tv,Psaltis:2013zja},
they are sufficient for a first data analysis
study, as shown through numerical simulations~\cite{Lo:2013ava}
and analytical estimates~\cite{Baubock:2015ixa}.

In scalar-tensor gravity, the model for the pulse profile depends not only on the
parameters
$\bm{\theta} = \{m, \theta_{\rm s}, \iota_{\rm o}, \Delta \theta, D, k_{\rm B}T, f, \textrm{EoS}\}$
that describe the source, its geometry and the observer in general relativity, but
also on the theory parameters $\alpha_0$ (we use $\log_{10}|\alpha_0|$ in practice)
and $\beta_0$.
In this parameter space, $m$ is the mass of the star, which given a specified
EoS determines its radius $R$, $D$ is the distance to the source,
$f$ is the star's rotation frequency,
$\theta_{\rm s}$ is the colatitude of the center of the hotpot and
$\iota_{\rm o}$ is the inclination of the line of sight, both angles
measured relative to the star's rotation axis.
Since we must first calculate neutron star models in scalar-tensor gravity,
instead of using $m$ we use the central energy density
$\varepsilon_{15} \equiv \varepsilon_{\rm c} \times 10^{-15}$ g/cm$^{-3}$
as our model parameter.
In summary, our full set of model parameters are
$\bm{\theta} = \{\varepsilon_{15}, \theta_{\rm s}, \iota_{\rm o}, \Delta \theta, D, k_{\rm B}T, f, \textrm{EoS}, \log_{10}|\alpha_0|, \beta_0 \}$.

%%%%%%%%%%%%%%%%%%%%%%%%%%%%%%%%%%%%%%%%%%%%%%%%%%%%%%%%%%%%%%%%%%%%%%
\section{Projected constraints from observations}
\label{sec:constraints}

Given the model described above, let us now roughly estimate the accuracy to which
scalar-tensor theories could be constrained given a NICER observation
that \textit{is consistent with general relativity}.
That is, we generate a synthetic signal (or injection) $\bf{d}$
using the Schwarzschild+Doppler model in general relativity, and we
then attempt to extract it and estimate its parameters with the
Just+Doppler model.

A detailed analysis of the full likelihood is important, but beyond the scope of this
work if one is concerned only with forecasting preliminary constraints. If the \textit{forecasted
constraints were to reveal that NICER cannot place interesting bounds on scalar-tensor
gravity, then a more detailed analysis would not be necessary}. However, in the opposite case,
one would conclude that NICER may be used as a laboratory to
test general relativity, and a more detailed data analysis study would be justified.
As we will show next, the latter is the case here.

The parameter space of the model is ten-dimensional and thus,
to arrive at a rough estimate on projected constraints, we
make a few simplifications that reduce this dimensionality.
We fix $\{\theta_{\rm s}, \iota_{\rm o}, \Delta \theta, D, k_{\rm B}T, f, \textrm{EoS} \}$
at their injected values (indicated by asterisks). For the first five, we set
$\{\theta_{\rm s}^{\ast}, \iota_{\rm o}^{\ast}, \Delta \theta^{\ast}, D^{\ast},
k_{\rm B}T^{\ast}\} = \{90^{\circ}, 90^{\circ}, 0.01^{\circ}, 600\,\textrm{pc}, 0.35\,\textrm{keV}\}$.
Our choices for $\theta_{\rm s}^{\ast}$ and $\iota_{\rm o}^{\ast}$ are motivated by the
results of~\cite{Lo:2013ava,Lo:2018hes}, which showed that this is the best-case orientation for
determining the mass and radius of the source, with a $5\%$ accuracy when the distance
to the star is also known (as assumed here).
The value of $\Delta\theta^{\ast}$ is chosen to enforce the point-like hot spot approximation,
while we chose typical illustrative values for $D^{\ast}$ and $k_{\rm B} T^{\ast}$.
Next, for $\textrm{EoS}^{\ast}$
we consider both a soft (SLy4) and a stiff (APR) EoS in order to
test the variability of our constraints with the EoS. With an EoS$^{\ast}$ chosen,
we choose $\varepsilon^{\ast}_{15}$ to yield a neutron star with $1.93$ M$_{\odot}$
as, e.g.~is the case of the neutron star in the PSR J1614--2230 binary~\cite{Fonseca:2016tux,Miller:2016kae}.
This choice is motivated by the observation that high-mass stars are more susceptible
to the presence of the scalar field than low-mass ones (cf.~Fig.~\ref{fig:mass_radius}).
We also consider two values of $f^{\ast}$, 200 and 600 Hz, the latter
to magnify the contributions of Doppler and relativistic aberration effects on the
pulse profile\footnote{For these large rotation frequencies one should
include the influence of stellar oblateness in the model~\cite{Miller:2014mca,Salmi:2018gsn} using
e.g.~the Oblate+Schwarzschild approximation~\cite{Cadeau:2006dc,Morsink:2007tv}. As
shown in these papers, when $\theta_{\rm s}$, $\iota_{\rm o}$ are close to the
equator, the effects of oblateness on the pulse profile are suppressed,and we expect
the same to be true in scalar-tensor gravity~\cite{Silva:2018yxz}.}~\cite{Poutanen:2006hw,Sotani:2018oad}.
Finally, to generate our general relativistic injection, we
set $\log_{10} |\alpha^{\ast}_0| = -5$ and
$\beta_{0}^{\ast} = 0$\footnote{As noted earlier, these stars have $Q \approx 10^{-5}$ and
are thus indistinguishable from neutron stars in general relativity as far as
NICER is concerned.}.

With these simplifications, our parameter space becomes three-dimensional,
spanning only the central energy density $\varepsilon_{15}$ and the
theory parameters $\log_{10}|\alpha_0|$ and $\beta_0$.
For $\varepsilon_{15}$, we assume uniform priors in the ranges
$\varepsilon_{15} \in [0.7, 3.0]$ (if EoS$^{\ast} = $ SLy4) and
$\varepsilon_{15} \in [0.5, 2.4]$ (if EoS$^{\ast} = $ APR), which
cover neutron star models with (roughly) minimum masses of 1~M\msun\, and
beyond the maximum mass for the corresponding EoS.
For $\beta_0$, we also use an uniform prior
$\beta_0 \in [-7, 0]$, which includes the regime where
spontaneous scalarization happens.
The lower-bound already violates the best bounds from binary-pulsar
observations by more than 1$\sigma$~\cite{Freire:2012mg,Shao:2017gwu,Anderson:2019eay},
whereas the upper bound is chosen to include the limit of general relativity.
Finally, for $\log_{10}|\alpha_0|$ we choose an uniform prior
$\log_{10}|\alpha_0| \in [-5, -2]$. The lower bound causes negligible
changes to the structure of neutron stars, whereas the upper bound
already violates the Cassini constraint by approximately one order of magnitude.

\begin{figure}[t]
\includegraphics[width=0.48\textwidth]{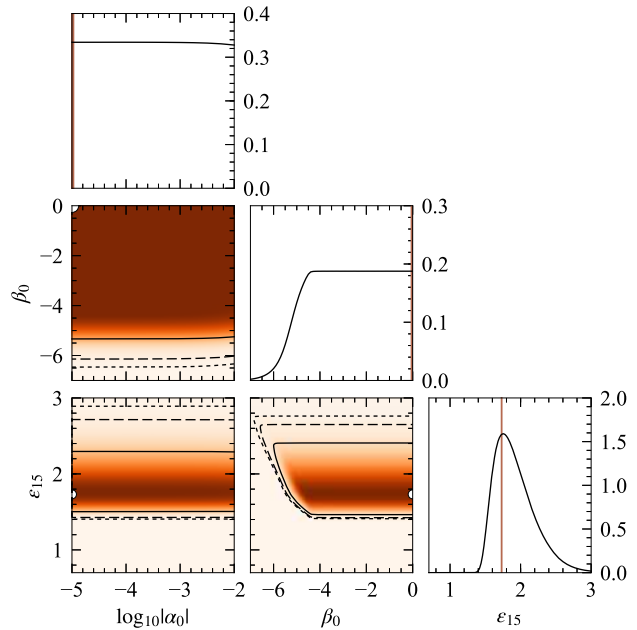}
\caption{Illustrative corner plot for the marginalized posterior distributions on the
    $\{\log_{10}|\alpha_0|, \beta_0, \varepsilon_{15}\}$ parameter space
    using EoS SLy4 and assuming
    a rotation frequency $f = 600$ Hz. In the color maps, the contours represent the $50\%$ (solid line),
    $90\%$ (dashed line) and $95\%$ (dotted line) credible regions. The (white) circles
    indicate the injection values $\{-5, 0, 1.731\}$. These are indicated by the
    vertical lines in the panels showing the marginal posteriors.
    The results are qualitatively the same with EoS APR, albeit the
    constraints on $|\alpha_0|$ and $\beta_0$ are weakened
    when $f$ is reduced and strengthened when we use this EoS
    (cf.~Fig.~\ref{fig:constraints}).
}
\label{fig:posteriors}
\end{figure}

Having reduced our parameter space and chosen our priors, we calculate
the best-fit parameters by minimizing the relative
chi-squared between the injection and the model pulse profiles, scanning the
parameter space using 16 phase stamps over the course of one stellar revolution.
The standard deviation of the distribution $\sigma_{\varepsilon_{15}}$
is modeled as in~\cite{Ayzenberg:2016ynm,Ayzenberg:2017ufk} and chosen
to capture the optimistic $5\%$ accuracy at which NICER can infer $m$ and $R$.
From the chi-squared, we calculate the likelihood function and from it
we obtain the marginalized posterior distributions
$p(\varepsilon_{15}|\bf{d})$, $p(\log_{10}|\alpha_0| | \bf{d})$ and $p(\beta_0|\bf{d})$.

Figure~\ref{fig:posteriors} summarizes our results
for $f^{\ast} = 600$ Hz and EoS$^{\ast} = $ SLy4.
The posterior of $\log_{10}|\alpha_0|$ is essentially
flat, so little information is gained relative to the uniform prior.
The posterior of $\beta_0$ is
more interesting, clearly exhibiting a sharp decay that starts near
the scalarization threshold at $\beta_0 \approx -4.35$.
When $\beta_0 \gtrsim -4.35$,
neutron stars are nearly identical to those of general relativity resulting in a
flat marginal posterior in this region.
However, when $\beta_0 \lesssim -4.35$, the conditions for
spontaneous scalarization are realized, and, as
neutron stars deviate more and more from their general relativity
cousins as $\beta_0 \to -7$, they result in poor models to recover the injection,
yielding an almost zero marginalized posterior when $\beta_0 = -7$.
The posterior for $\varepsilon_{15}$ peaks within $\lesssim 1\%$ from the injected value $\varepsilon^{\ast}_{15}$,
with a slow decay as $\varepsilon_{15} \to 3$. This is because
stars with the SLy4 EoS have a small variation in their radii and masses
past the maximum mass ($\approx 2$ M\msun), which is close to the injected value
($1.93$ M\msun).

\begin{figure}[t!]
\includegraphics[width=0.48\textwidth]{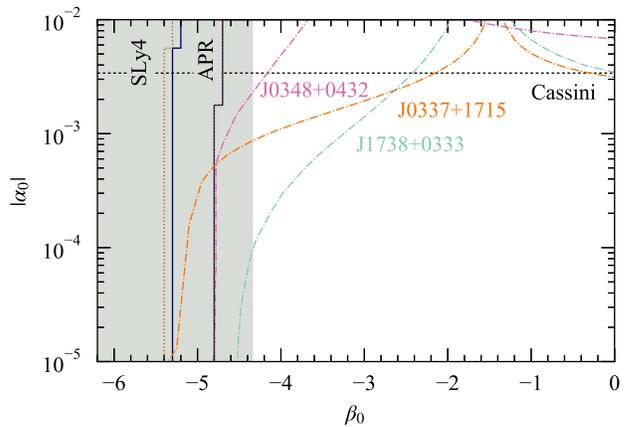}
\caption{Projected $1 \sigma$-constraints on ($\alpha_0,\, \beta_0$)
from pulse profile observations. In this plane, general relativity
($|\alpha_0| = 0,\, \beta_0 = 0$) is pushed infinitely down the ordinate.
The shaded region covers the portion of the parameter space in which spontaneous
scalarization happens.
The solid (dashed) lines delimits the 68$\%$ credible regions for stars
rotating with $f=600$ (200) Hz, and different equations of state.
For comparison, we also included the 68$\%$ credible regions from the Cassini
mission~\cite{Bertotti:2003rm} (dotted line) and the binary-pulsar
systems J1738-0333, J0348-0432~\cite{Anderson:2019eay} and the
hierarchical triple system PSR J0337-1715~\cite{Archibald:2018oxs} (dot-dashed lines).
The left-most part of our constraints lays on $\beta_0 = -4.8$
(for the APR EoS), indicating that pulse profile observations
have the potential to marginally exclude strong-gravity phenomena as
spontaneous scalarization, which in this particular theory requires
$\beta_0 \lesssim - 4.35$.}
\label{fig:constraints}
\end{figure}

The conclusions above are robust to changes in the injected rotation frequency
and the EoS. When we set $f^{\ast}$ to 200 Hz and the EoS$^{\ast}$
to APR, the likelihood surface does not change considerably, although
the constraints on $\beta$ are more stringent than when injecting SLy4,
as summarized in Fig.~\ref{fig:constraints}. The latter shows the
68$\%$ credible regions in the $(|\alpha_0|,\beta_0)$-plane for both EoS injections,
which ought to be compared with the ($1\sigma$) constraints from Cassini and from
the absence of scalar dipole radiation from the PSR J1738-0333\cite{Freire:2012mg,Antoniadis:2012vy},
PSR J0348-0432~\cite{Kramer:2006nb,2009CQGra..26g3001K} binaries~\cite{Anderson:2019eay}
and the hierarchical triple system PSR J0348-1715~\cite{Archibald:2018oxs}.
The contours reveal that for stiff equations of state (such as APR) the
constraints on scalar-tensor gravity (in the $|\alpha_0| \to 0$
portion of the parameter space) from NICER observations can, in principle,
be as strong as those obtained by binary pulsar observations\footnote{
A note of caution is necessary when comparing the constraints obtained here
and Refs.~\cite{Anderson:2019eay,Archibald:2018oxs} shown in Fig.~\ref{fig:constraints}.
The constraints on $\beta_0$ depend critically on the choice of
the prior range and a na\"ive comparison between the three works is strictly
not correct. Moreover, the curves shown from~\cite{Anderson:2019eay} refer
specifically to EoS SLy4, whereas the EoS used in~\cite{Archibald:2018oxs} (labeled ``0.2'' in~\cite{1981A&A...102..299H})
is considerably stiff, with a 2.6 M$_\odot$ maximum mass in general relativity.
}.

Let us further compare and contrast the qualitative behavior of the constraint contours of Fig.~\ref{fig:constraints}.
As mentioned previously, $\beta_0$ is unconstrained from Solar System experiments, and
therefore, the Cassini contour is constant at a fixed $|\alpha_0|$.
For binary pulsars, when $\beta_0 > -4.35$ neutron stars have small $Q\approx|\alpha_0|$,
and the modification to the binary's orbital time decay $\dot{P}_{b}$ scales with $|\alpha_0|^2$.
But this modification is enhanced by two powers of the orbital velocity, which is of ${\cal{O}}(10^{-2}-10^{-3})$,
because the scalar field generates dipole radiation, instead of the quadrupole radiation of general relativity.
The combination of these effects results in a modification to $\dot{P}_{b}$ that can be constrained
even for values of $(\alpha_{0},\beta_{0})$ beyond the threshold of scalarization and below the
Cassini bound.
Finally, in the NICER case, the modifications to the spacetime are controlled by $Q$, which is
small and roughly independent of $\alpha_{0}$ inside our prior range and outside the threshold of scalarization
(i.e. outside the shaded region in Fig.~\ref{fig:constraints}). When scalarization does occur, however, $Q$ acquires
a non-perturbative enhancement, that leads to a large deviation from general relativity, and thus to nearly vertical
constraint contours.

%%%%%%%%%%%%%%%%%%%%%%%%%%%%%%%%%%%%%%%%%%%%%%%%%%%%%%%%%%%%%%%%%%%%%%
\section{Conclusions}
NICER allows for the observation of x-ray pulse profiles from rotating
neutron stars with unparalleled timing resolution. As the signal
travels from the star to the detector, it probes the spacetime in the
star's vicinity, opening the possibility of glimpsing (or
constraining) deviations from general relativity in the strong-gravity
regime.
Using the tools developed in~\cite{Silva:2018yxz}, we have
explored whether such observations can place constraints on a concrete
scalar-tensor gravity model which admits large deviations from
general relativity \textit{only} in strong-gravity environments~\cite{Damour:1993hw,Damour:1996ke}.

As a proof-of-principle, we carried out a restricted likelihood analysis
that suggests that (in principle) observations carried by NICER can constrain the
parameter space of scalar-tensor gravity, severely restricting the domain in
which spontaneous scalarization occurs.
These projections are clearly preliminary but they do demonstrate,
for the first time, the potential of including pulse
profile observations as a new tool in the relativists' arsenal to perform strong-gravity tests.
This is of interest not just for the ongoing NICER mission, but also for planned,
future missions, such as the \textit{Enhanced X-ray Timing and Polarimetry}
mission~\cite{intZand:2018yct}.
Given these preliminary projections, \textit{there is now strong justification
to carry out a more detailed data analysis study that includes
a more careful modeling of the instruments's response, as well as the
covariances between all parameters in the model
to determine how much degeneracies impact our ability to test
general relativity.}

The encouraging results found here motivate the consideration of other questions.
For instance, one could use both a general relativistic
and a scalar-tensor model to analyze real NICER data. Using Bayesian analysis,
one could then perform a model selection study to determine which theory is better
supported by the data through the Bayes factor.
Another important question is that of fundamental bias~\cite{Yunes:2009ke} in the
parameter estimation of the properties of neutron stars.
Pulse profile models assume \textit{a priori} that general relativity is correct,
which may bias parameter estimation, if the true underlying
theory is not Einstein's.
As a consequence, the mass and radius of the source could be systematically biased with immediate
implications for EoS inference.
This question could be investigated through a Bayesian analysis that uses
a scalar-tensor theory pulse profile injection, but general relativity to model the data.

Still in the realm of synthetic data, a number of improvements can also be made.
First, we could include beaming of the emitted radiation
caused by Thomson scattering in the stellar atmosphere~\cite{Lo:2013ava,Miller:2014mca,Miller:2016kae,Salmi:2018gsn}.
Second, as radiation travels towards the observer, it interacts
with the star's accretion disk~\cite{Poutanen:2003yd} and the
interstellar medium~\cite{Miller:2016kae,Salmi:2018gsn}.
Third, one could include background
radiation~\cite{Ozel:2015ykl,Lo:2013ava,Miller:2016kae} and
generate synthetic data that is as realistic as possible.
This data would then have to be convolved with the detector's response matrix,
and combined with a Poisson sampling of the modeled pulse
profile to determine the chi-squared through a Poisson likelihood~\cite{Miller:2016kae}.
With these ingredients taken into account, it would be interesting
to verify whether the constraints on scalar-tensor gravity obtained here are
\textit{robust} to the assumptions we have made throughout the text.
Regardless of the simplifications made here, our work nonetheless puts forward the enticing
potential of probing strong-gravity with pulse profile observations.

\appendix
\section{Data analysis}

In this appendix we present a brief overview of the methods used to
obtain the projected constraints obtained in Sec.~\ref{sec:constraints}.
%
% Our approach follows closely that used in~\cite{Ayzenberg:2016ynm,Ayzenberg:2017ufk}.

We call the signal measured during an observation the
{\it synthetic injected signal}, or (for brevity)
{\it the injection} $F_{\rm inj} (\bm{\theta}^{\ast})$.
The pulse profile that we use to extract and characterize this observed pulse profile
is referred to as {\it the model} $F_{\rm mod}(\bm{\theta})$.
Both pulse profiles can be calculated following~\cite{Silva:2018yxz} once all
parameters
$\bm{\theta} = \{\varepsilon_{15}, \theta_{\rm s}, \iota_{\rm o}, \Delta \theta, D, k_{\rm B}T, f, \textrm{EoS}, \log_{10}|\alpha_0|, \beta_0 \}$
have been specified.

As discussed in the main text, we consider the reduced model parameter space
obtained by fixing
$\bm{\theta}_{\rm fix} = \{\theta_{\rm s}, \iota_{\rm o}, \Delta \theta, D, k_{\rm B}T, f, \textrm{EoS}\}$
to the injected values, leaving as variable model parameters
$\bm{\theta}_{\rm var} = \{\varepsilon_{15}, \log_{10} |\alpha_0|, \beta_0 \}$

We calculate the best-fit parameter values by minimizing the
reduced chi-squared $\chi^2_{\rm red}$ between the injection and the model pulse profiles,
sampling over the model's variable parameters ${\bm \theta}_{\rm var}$.
The reduced chi-squared is defined as
\begin{equation}
\chi_{\rm red}^2 \equiv
% \frac{\chi^2}{N}
% \nonumber \\
% &=
\frac{1}{N}\sum_{i = 1}^{N}
\left[
\frac{F_{\rm mod}(\phi_i, {\bm \theta}_{\rm fix}, {\bm \theta}_{\rm var}) - F_{\rm inj}(\phi_i, {\bm \theta}_{\rm fix}, {\bm \theta}^{\ast}_{\rm var})}{\sigma(\phi_i)}
\right]^2\,,
\label{eq:red_chi}
\end{equation}
where the summation is over $N$ time-stamps during the course of one revolution
of the star. We normalize the phase over a revolution such that $\phi_i \in [0, 1]$,
and use $N = 16$ time stamps.
The standard deviation of the distribution $\sigma$ is modeled as
\begin{equation}
\sigma(\phi_i) =
\sigma_{\varepsilon_{15}}(\phi_i) +
\sigma_{\log_{10}|\alpha_0|}(\phi_i) +
\sigma_{\beta_0}(\phi_i)\,,
\end{equation}
where $\sigma_{\varepsilon_{15}}$, $\sigma_{\log_{10}|\alpha_0|}$ and $\sigma_{\beta_0}$ are the
standard deviations on each injected parameter
$\bm{\theta}^{\ast}_{\rm var} = \{\varepsilon_{15}^{\ast}, \log_{10} |\alpha_0^{\ast}|, \beta_0^{\ast} \}$.

Since our injection is assumed to be consistent with general relativity, we set
$\sigma_{\log_{10}|\alpha_0|} = \sigma_{\beta_0} = 0$,
while the standard deviation $\sigma_{\varepsilon_{15}}$ is calculated by~\cite{Ayzenberg:2016ynm,Ayzenberg:2017ufk}
\begin{align*}
    \sigma_{\varepsilon_{15}}(\phi_i) &= \frac{1}{2}
\left\vert
F_{\rm inj}(\phi_i, {\bm \theta}_{\rm fix}, \{\varepsilon_{15}^{\ast} + \delta \varepsilon_{15}^{+}, \log_{10} |\alpha_0^{\ast}|, \beta_0^{\ast} \})  \right.
\nonumber \\
&\quad \left.
    - \,\, F_{\rm inj}(\phi_i, {\bm \theta}_{\rm fix}, \{\varepsilon_{15}^{\ast} - \delta \varepsilon_{15}^{-}, \log_{10} |\alpha_0^{\ast}|, \beta_0^{\ast} \})
\right\vert\,,
\end{align*}
where (as explained in the main text)
$\log_{10} |\alpha_0^{\ast}| = -5$ and $\beta_0^{\ast} = 0$,
while $\delta \varepsilon_{15}^{\pm}$ remain to be specified.
To obtain them, we start by assuming that $m^{\ast} = 1.93$ M$_\odot$ as
discussed in the main text.
Then, given an EoS, we calculate the
central energy density $\varepsilon_{15}^{\ast}$ for which this mass is
obtained and record the corresponding radius $R^{\ast}$.
Next, we draw
a box in the mass-radius plane (in general relativity) centered at $(m^{\ast}, R^{\ast})$
with width spanning $R^{\ast}(1 \pm 0.05)$ and height spanning $m^{\ast}(1 \pm 0.05)$.
The mass-radius curve intersects this box twice and these two points then determine
$\delta \varepsilon_{15}^{\pm}$.

Once the reduced chi-squared is obtained, we calculate the likelihood
\begin{equation}
    L({\bm \theta}_{\rm var}) = \exp\left(- \chi_{\rm red}^2 / 2\right)\,.
\end{equation}
When this is evaluated over the whole prior domain, we obtain the likelihood distribution
shown in the off-diagonal panels of Fig.~\ref{fig:posteriors}. Upon marginalization of the likelihood,
we obtain the diagonal panels of the same figure.

To obtain the constraints shown in Fig.~\ref{fig:constraints}, we first sort
our likelihood data in decreasing order of values of $L(\bm{\theta}_{\rm var})$.
Then, we sum all the values of
$L$ (and record the corresponding $\bm{\theta}_{\rm var}$) until $68\%$ of the
sum over \textit{all} $L$-values is reached.
The values of $\log_{10}|\alpha_0|$ and $\beta_0$ obtained from this procedure
result in the dotted and solid lines in Fig.~\ref{fig:constraints}.
%

%%%%%%%%%%%%%%%%%%%%%%%%%%%%%%%%%%%%%%%%%%%%%%%%%%%%%%%%%%%%%%%%%%%%%%
\section*{Acknowledgments}
We thank Anne Archibald, Justin Alsing, Alejandro C\'ardenas-Avenda\~no,
Neil Cornish, Cole Miller, Sharon Morsink, George Pappas and
Hajime Sotani for valuable discussions.
We also thank David Anderson for calculating and sharing with us the
binary pulsar constraints.
Finally, we thank the anonymous referees for pertinent comments on
our work.
This work was supported by NASA grants NNX16AB98G and 80NSSC17M0041.
Computational efforts were performed on the Hyalite High Performance
Computing System, operated and supported by  University Information Technology
Research Cyberinfrastructure at Montana State University.
%%%%%%%%%%%%%%%%%%%%%%%%%%%%%%%%%%%%%%%%%%%%%%%%%%%%%%%%%%%%%%%%%%%%%%s
\bibliographystyle{apsrev4-1}
\bibliography{biblio}

\end{document}